\documentclass{elsart3p}
\usepackage{graphicx}
\usepackage{amssymb}
\journal{Physics Letters A}
\begin{document}
\begin{frontmatter}
\title{X-ray and neutron diffraction studies of coupled structural phase
transitions in DyBaCo$_{2}$O$_{5.5}$}
\author[PNPI]{Yu.P.~Chernenkov\corauthref{cor}},
\corauth[cor]{Corresponding author.}
\ead{yucher@pnpi.spb.ru}
\author[PNPI]{V.P.~Plakhty},
\author[CEA]{A.G.~Gukasov},
\author[Minsk]{S.N.~Barilo},
\author[Minsk]{S.V.~Shiryaev},
\author[Minsk]{G.L.~Bychkov},
\author[MPI]{V.~Hinkov},
\author[PNPI]{V.I.~Fedorov},
\author[PNPI]{V.A.~Chekanov}
\address[PNPI]{Petersburg Nuclear Physics Institute, RAS, RU-188300 Gatchina, St.Petersburg, Russia}
\address[CEA]{Laboratoire Leon Brillouin (CEA-CNRS), CE Saclay, FR-91191 Gif-sur-Yvette, France}
\address[Minsk]{Institute of Solid State and Semiconductors, BAS, 17 P.Brovka st., BY-220072 Minsk, Belarus}
\address[MPI]{Max-Plank-Institut f\"{u}r Festk\"{o}rperforschung, DE-70569 Stuttgart, Germany}
\begin{abstract}
A structural transition at $T\approx 322$ K from the $Pmmm$ to $Pmma$ phase is found to
coincide with an anomaly of resistivity. Another structural phase transition
doubling the lattice parameter $c$, which has been postulated earlier to
accompany a low-temperature magnetic transition in TbBaCo$_{2}$O$_{5.5}$, is
observed in a single crystal DbBaCo$_{2}$O$_{5.5}$ by means of the X-ray and
neutron diffraction. The low temperature phase does not belong to the space
group $Pcca$ that has been chosen earlier as the highest subgroup of the $Pmma$. The
transition is of the first order with the temperature hysteresis, between
$T\approx 100$ and $T\approx 200$ K, which probably explains anomalous
magnetic properties in this temperature range.
\end{abstract}

\end{frontmatter}
\section{Introduction}
The rare-earth ($R)$ oxides $R$BaCo$_{2}$O$_{5+\delta }$ have a double-perovskite
structure, with all the cobalt ions formal oxidation being 3+ at \textit{$\delta $}~=~0.5. A
competition between the intra-atomic exchange and the crystal field results
in three possible spin states of the Co$^{3+}$ ions: the low-spin (LS,
$t_{2g}^6 e_g^0 $, $S$~=~0), the intermediate-spin (IS, $t_{2g}^5 e_g^1 $,
$S$~=~1), and the high-spin (HS, $t_{2g}^4 e_g^2 $, $S$~=~2) \cite{1}. The small energy
differences between these states as well as a twofold degeneracy of the IS
electronic configuration leads to the quite unusual magnetic properties
including coupled phase transitions
(structural/~spin-order/~spin-state-order/~orbital-order) and, probably as a
consequence, the giant magnetoresistance. These properties best of all
develop themselves in the crystals with well ordered alternating planes
$(0,1,0)$ at $y=0$ and $y=1/2$ of the Co$^{3+}$ ions in the oxygen pyramids
and octahedra, respectively, as shown schematically in Fig.~1 by the grey
and white circles. We consider in this paper the only case of $\delta\approx
0.5$ that corresponds to the superstructure $a_{2}\approx a_{p}$,
$b_{2}\approx 2a_{p}$, $c_{2}\approx  2a_{p}$, where $a_{p}$ is a
perovskite lattice parameter (Fig.~1). If $\delta$  differs considerably from 0.5, an
additional degree of freedom, the charge, should be added, which may be
ordered giving unusual superstructures.

The regular study, e.g. \cite{2}, of the crystal lattice dependence on the oxygen
content for a number of slowly
cooled $0.17<\delta<0.79$ and rapidly quenched $0.23<\delta<0.71$
samples PrBaCo$_{2}$O$_{5+\delta}$ has shown that for the samples, which
are expected to be most disordered, the unit cell is close to a tetragonal
one with the lattice parameters $a_{1}\approx a_{p},b_{1}\approx
a_{p},c_{1}\approx 2a_{p}$. Various superstructures, mainly with
short-range ordering, were observed \cite{3} in the family $R$~=~Pr~--~Ho, with
$\delta$ being varied from 0.3 to 0.7. However, even at the ideal oxygen content,
$\delta=0.5$, the ordering of the apical oxygen atoms $O_1,{O}'_1,O_2
,O_3$ shown in Fig.~1 by the black circles controls the ordering degree
of the alternating octahedral and pyramidal Co$^{3+}$ ions. As an example
\cite{4}, we give the apical oxygen distribution in two single crystals
Tb$_{0.9}$Dy$_{0.1}$BaCo$_{2}$O$_{5+\delta}$: 1.~$n(O_{1})=1.00(1)$,~$n({O}'_1)=0.09(2)$,
~$n(O_{2})=1.00(2)$,~$n(O_{3})=1.00(1)$, with $\delta=0.55(3$); 2.~$n(O_{1})=1.00(1)$,
~$n({O}'_1)=0.33(2$),~$n(O_{2})=0.99(3)$,~$n(O_{3})=0.86(5)$ with $\delta=0.54(3)$.
Obviously, the apical oxygen ordering should be verified in addition to the
total oxygen content.

The materials with $a_{2}\approx a_{p}, b_{2}\approx 2a_{p}, c_{2}\approx2a_{p}$
are commonly described by the space group $Pmmm$ above
the room temperature. However, a series of very weak superstructure X-ray
reflections from a single crystal GdBaCo$_{2}$O$_{5.47(2)}$ observed \cite{5}
below $T_{S1}=341.5(2)$ K give an evidence of the second order structural
phase transition. The unit cell below the $T_{S1}$ has the lattice parameters
$a_{3}\approx 2a_{p}, b_{3}\approx ~2a_{p}, c_{3}\approx2a_{p}$,
 and the systematic extinction of the superstructure reflections
corresponds to the space group $Pmma$. Atomic displacements found from the
intensities of the superstructure reflections indicate an orbital/spin-state
ordering. The neutron powder diffraction patterns of
TbBaCo$_{2}$O$_{5.53(1)}$ have been successfully treated in the frame of
this group \cite{6}. Although small atomic displacements cannot influence a
powder diffraction pattern, the distribution of the equivalent Co$^{3+}$
ions over octahedral and pyramidal sites is completely different in
comparison with the $Pmmm$ group. Among the pyramidal cobalt ions, the ions 1 and
4 as well as 2 and 3 are equivalent in the $Pmma$, while all four ions are
equivalent in the $Pmmm$. The octahedral ions 5 and 8 as well as 6 and 7
yield a similar chess-board-like order (Fig.~1). This difference is very
important for the magnetic/spin-state/orbital ordering, and the group $Pmma$ has
been later verified \cite{4} with a twin-free single crystal of
Tb$_{0.9}$Dy$_{0.1}$BaCo$_{2}$O$_{5.54(3)}$, the material that is expected
to have the properties similar to the Tb ones studied in \cite{6}. The first
objective of our work is to check whether this phase transition is typical
for the $R$BaCo$_{2}$O$_{5.5}$ materials.
\begin{figure}
\begin{center}
\includegraphics*[width=7.5cm]{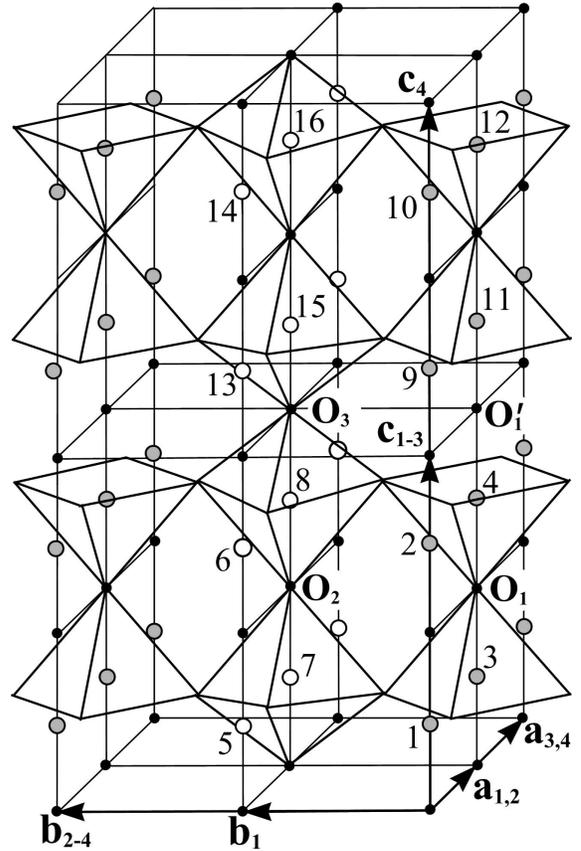}
\end{center}
\label{fig.1}
\caption{Four unit cells, observed in $R$BaCo$_{2}$O$_{5+\delta }$ at different
oxygen content, a degree of oxygen ordering and temperature. Cobalt ions in
the pyramidal and octahedral sublattices are shown by the grey and white
circles, respectively. Oxygen atoms in the apical positions are described by
black circles. The other oxygen atoms are in the corners of the coordinating
polyhedra. The $R$ and Ba atoms are omitted for simplicity.}
\end{figure}

The value of $T_{S1}$ measured for GdBaCo$_{2}$O$_{5.47(2)}$ is close to the
metal-insulator transition $T_{MI}=359$ K for the same oxygen content \cite{7}.
Since the transport properties should depend on the orbital ordering \cite{8},
the second objective is to verify whether these critical temperatures
coincide when measured for the same crystal in identical temperature
conditions.

To explain the magnetic structure at $T=100$ K
of TbBaCo$_{2}$O$_{5.53(1)}$ we have postulated \cite{6} a structural phase transition at about 170~K
to the phase with the lattice parameters $a_{4}\approx 2a_{p},
b_{4}\approx 2a_{p}, c_{4}\approx 4a_{p}$, as shown in Fig.~1.
Since the experiment was made with a ceramic sample, and no very weak
superstructure reflections could be observed for selecting the space group
by their extinction, we have suggested a highest subgroup of the $Pmma$ with the
wave-vector $\textbf{k} = \textbf{c}_3^\ast/2$, which is $Pcca$, for the space group of the
hypothetical new phase. Our third, the main objective, is to look for the
low-temperature structural phase transition, to investigate its character
and to verify whether the systematic extinction of the X-ray superstructure
reflections yields the $Pcca$ group suggested earlier.

\section{Experimental methods}
Single crystals of DyBaCo$_{2}$O$_{5+\delta }$ were grown from an
overstoichiometric fluxed melt which was prepared in accordance with optimal
composition in the DyO$_{1.5}$~--~Ba$_{2}$Co$_{3}$O$_{y}$ cross section of
the Gibbs triangle \cite{4}. Relatively large (up to $5\times 6\times
0.5~mm^{3}$) single crystals of rectangular shape were grown in a furnace
with a vertical chamber supplied by Cr-La resistive heaters. A batch of
Dy$_{2}$O$_{3}$ was put as feeder on the bottom of a 100 cm$^{3}$ magnesia
crucible to keep the flux melt saturated at temperature of about 1200\r{}C
under positive temperature gradient 1--3\r{}C and to grow a limited
number of nuclei for about three weeks of isothermal crystal growth process.
The as-grown single crystals were annealed in flowing oxygen for about three
days. The cooling with rate of 10\r{}C/h from 660\r{}C down to 330\r{}C
and then with rate 20\r{}C/h down to room temperature was used to get
$\delta=0.50(2)$ according to iodometric titration. One of these oxygenated and
twinned crystals DyBaCo$_{2}$O$_{5.5}$ was used in the X-ray diffraction and
resistivity measurements. The total oxygen content and the oxygen ordering
were verified by means of X-ray diffraction as explained in the
introduction. The X-ray value of $\delta$ coincided with that obtained by the
iodometric titration in the limits of a standard deviation. As to the oxygen
distribution over the apical sites, a small amount (7~--~10~{\%}) of oxygen
was always present in the site ${O}'_1 $ that should be vacant in the ideal
case.

One of the crystals was cut in a rectangular shape and polished. The edges
of the crystal were kept parallel to the crystallographic axes with an
estimated error less than 5 degree of arc. In order to remove twins the
crystal was kept for 12~h under pressure of about 0.15~--~0.20~GPa in
flowing oxygen at 300~--~350\r{}C and then quickly cooled to room
temperature. We have applied pressure along the longest edge of the
rectangular sample, and this direction corresponds to the smallest
crystallographic \textbf{a}-axis in the resulting twin-free structure of the
sample. The polarized-light microscope image and the susceptibility
characterization have indicated less then 4{\%} fraction of the sample
survived as \textbf{b}-axis oriented after such a procedure.

The in-plane resistivity $\rho$ of a twinned crystal was measured using
a standard dc four-probe method. Two current wires, as well as two voltage ones in
between, were connected with the room drying silver paste to one natural
crystal face. The face polishing could easily destroy the crystal that had
visible microscopic cracks. We were aware of the negative factors, like the
contact resistance and, in particular important, additional effects of the
surface conductivity. However, our aim was to search for an anomaly in the
temperature dependence of resistivity, and possible artifacts were not that
dangerous as in quantitative measurements of $\rho(T)$ itself. On contrary, we paid
attention to the crystal temperature. The temperature difference with the
X-ray experiment including its stability was estimated as about 2~K.

It is commonly believed that the structural changes due to the oxygen atoms
should be investigated by means of neutron diffraction. This is obvious when
taking into account the scattering ability of oxygen in comparison with
heavy atoms. In our case, the crystals were too small for investigation of
the superstructures due to the oxygen ordering that produce the Bragg
reflections a few orders of magnitude weaker than the basic ones. We
combined these two methods in our work. The nuclear and magnetic structure
as well as the magnetization density were investigated with a twin-free
crystal of $4\times 3\times 0.4~mm^{3}$ on 5C1 and 6T2
diffractometers at the ORPH\'{E}E reactor of the Laboratoire Leon Brillouin
(CE~Saclay, France). Here we present only the temperature dependence of some
Bragg reflections, mainly of the magnetic origin, to be compared with the
X-ray and resistivity data. The other results will be published elsewhere.
For historical reasons, the neutron studies were made at the beginning, and,
because of the cobalt activation by neutrons, we were left for the X-ray and
resistivity measurements with a twinned crystal that was checked to have the
total oxygen content and distribution over the apical positions as in the
first one, before the detwinning procedure. As explained in the next
section, a conclusion on the low-temperature crystal symmetry can be made
even with a twinned crystal. The crystal had thickness of about 0.2~mm, and
was enough transparent for the Mo~$K_{\alpha }$ radiation $\lambda=0.712 \textrm{\AA}$.
As usually \cite{5}, the voltage on a 2~kW X-ray tube was set lower then the
high-energy edge of the white spectrum for avoiding any $\lambda/2$ contaminations in
the primary beam after a PG monochromator.

\section{Results and discussions}
The temperature dependence of the peak intensity for the superstructure
reflection $(3,1,1)+(1,3,1)$, as well as the basic one
$(6,2,2)+(2,6,2)$ from a twinned crystal is displayed in Fig.~2(a), with
the indices being given for the unit cell with parameters $a_{3}$~$\approx
$~2$a_{p}$, $b_{3}$~$\approx $~2$a_{p}$, $c_{3}$~$\approx $~2$a_{p}$. A
continuous structural transition is definitely seen at $T_{S1}$~$\approx
$~323~K, and the temperature variation of both reflections is identical, as
shown by a solid line in Fig.~2(a), where two intensities are reduced to one
scale by their average value in the temperature range 293~--~321~K.
For comparison, a similar transition in GdBaCo$_{2}$O$_{5.47(2)}$ has been
observed at $T_{S1}$~$\approx $~341.5(7)~K. The X-ray data obtained earlier
\cite{4} for a twin-free crystal Tb$_{0.9}$Dy$_{0.1}$BaCo$_{2}$O$_{5.54(3)}$ has
given unambiguous evidence that at this transition the symmetry changes from
$Pmmm$ with the unit cell $a_{2}\approx a_{p}, b_{2}\approx 2a_{p},
c_{2}\approx 2a_{p}$ in the high-temperature phase to $Pmma$ with
$a_{3}\approx 2a_{p}, b_{3}\approx 2a_{p}, c_{3}\approx2a_{p}$.

We may conclude that this phase transition is a common
phenomenon, at least for the $R$BaCo$_{2}$O$_{5.5}$ materials with $R$ ions from
the middle of the rare-earth period. It has been suggested \cite{5,6} that the
structural changes are driven by the orbital/spin-state ordering of the
Co$^{3+}$ ions, which may be expected to influence the transport properties.
Possible mechanisms, like spin blockade, are suggested in \cite{8,9,10}. In the
case of DyBaCo$_{2}$O$_{5.5}$, the resistivity \textit{$\rho $}($T)$ starts to increase steeper
below $T_{S1}$, as seen in Fig.~2(b). It would be worthwhile to check the
coincidence of the two temperatures, $T_{S1}$ and $T_{MI}$, for the other
materials where this structural phase transition is observed.
\begin{figure}
\vspace{0.5truecm}
\begin{center}
\includegraphics*[width=7.5cm]{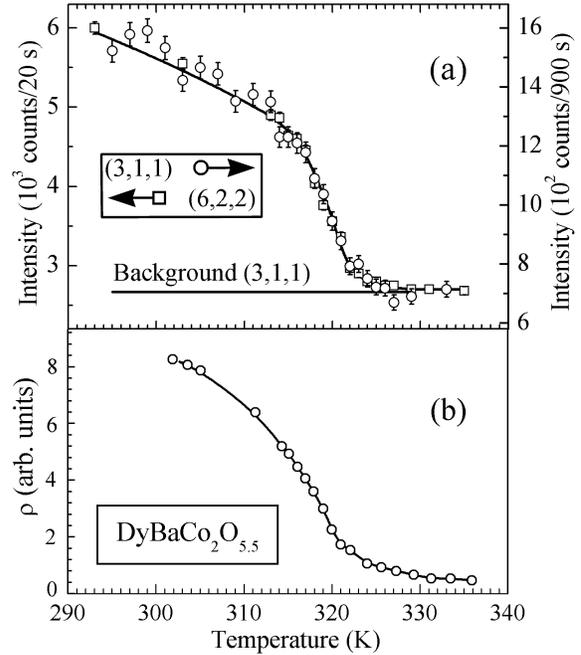}
\end{center}
\label{fig.2}
\caption{Variation with temperature of the peak intensity for the
superstructure $(3,1,1)+(1,3,1)$ and basic $(6,2,2)+(2,6,2)$
reflections (a) and of the in-plane resistivity}
\end{figure}

The intensities of the neutron reflections $(1,1,2)$, $(0,0,1)$
and $(1,1,1)$ (for $c_4=2c_3$) from a twin-free crystal
DyBaCo$_{2}$O$_{5.5}$ shown in Fig.~3(a) are mainly due to magnetic
scattering as well as those for the Tb material \cite{6}. Two of them with
$l$~--~odd indicate a spin antitranslation \cite{11} ${t}'=c_3 =c_4 /2$, while
$(1,1,2)$ gives an evidence of a translation $t=c_3 =c_4 /2$. To resolve the
issue, we have made an assumption \cite{6} that the low-temperature magnetic
transition in the Tb material coincides with a structural one that double
\textbf{c} axis: $c_{4}=2c_{3}$. Unlike the Tb material, reflection
$(1,1,1)$ is absent in the case of the DyBaCo$_{2}$O$_{5.5}$ below 200~K.
However, the other reflection $(0,0,1)$ with $l$~--~odd still coexists with
$(1,1,2)$, and the magnetic transition at 200~K should be accompanied by a
structural one as well. Intensity temperature dependence of the
superstructure X-ray reflection $(3,0,1)+(0,3,1)$ from the twinned
crystal DyBaCo$_{2}$O$_{5.5}$ is shown in Fig.~3(b).
\begin{figure}
\vspace{0.5truecm}
\begin{center}
\includegraphics*[width=7.5cm]{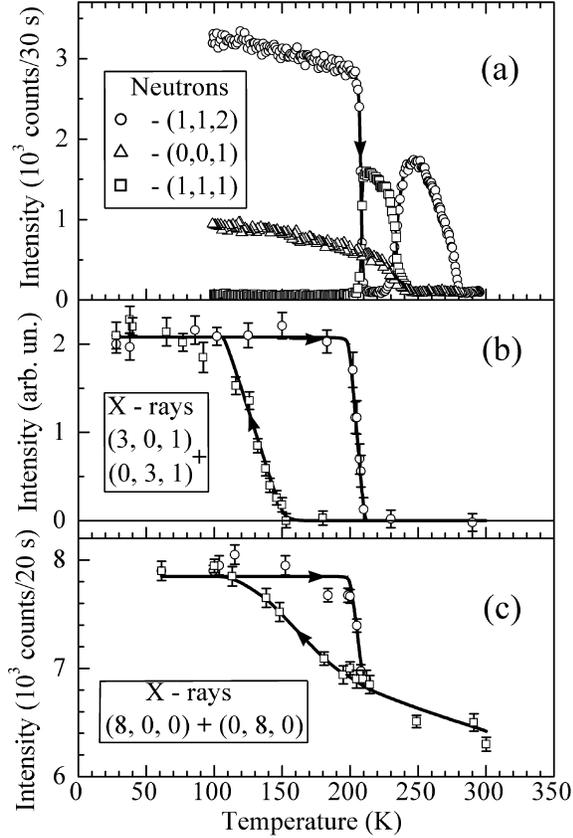}
\end{center}
\label{fig.3}
\caption{Intensity temperature dependence of three neutron Bragg reflections
from a twin-free crystal at heating (a) and X-ray reflections,
superstructure $(3,0,1)+(0,3,1)$ and basic $(8,0,0)+(0,8,0)$ from a
twinned crystal DyBaCo$_{2}$O$_{5.5}$ (b) and (c).}
\end{figure}
Corresponding rocking
curves are given in Fig.~4. One can definitely see a first order structural
transition, that begins at $T=200$ K and ends at $T=210$ K, when the crystal
is heating. On the crystal cooling this transition begins at $T=150$ K and
ends at $T=110$ K. The variation of the basic reflection
$(8,0,0)+(0,8,0)$ shown in Fig.~3(c) is similar to that of the
superstructure pair $(3,0,1)+(0,3,1)$. In the range of this wide
temperature hysteresis, the unusual magnetic properties of field-cooled and
zero-field-cooled crystals in the temperature range $120<T<210$ K
\cite{12,13}, probably, may be due to coexistence of two phases with different
crystal and magnetic structures.

In spite of the twins, it is definitively proved that $Pcca$ is not a correct
space group for the low-temperature phase. Both reflections, $(3,0,1)$ and
$(0,3,1)$, are forbidden in this group \cite{14}. A systematic inspection of the
reflections $(h,0,l)+(0,k,l)$ has shown that some of them forbidden by
symmetry, namely with $h$ and $l$ odd, are observed in contradiction with the
extinction law for the space group $Pcca$. We should remind that this group was
postulated as a highest subgroup \cite{15} of the $Pmma$ with the wave-vector
$\textbf{k}=\textbf{c}_3^\ast /2$. Strictly speaking, this argument can be exclusively used
for the second order phase transitions. Having the first order transition,
we can but try different groups with the proper extinction laws.
\begin{figure}
\vspace{0.5truecm}
\begin{center}
\includegraphics*[width=7.5cm]{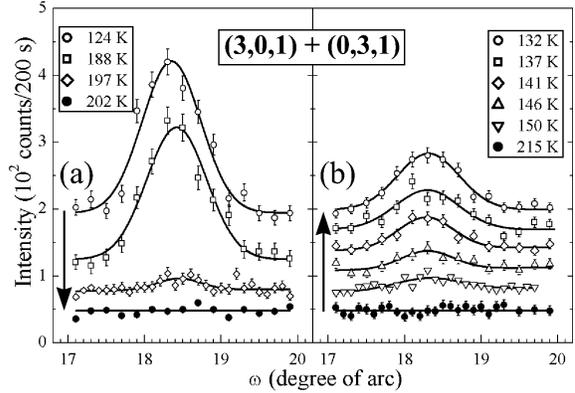}
\end{center}
\label{fig.4}
\caption{Rocking curves of the X-ray superstructure reflection
$(3,0,1)+(0,3,1)$ from a twinned crystal DyBaCo$_{2}$O$_{5.5}$ at
different temperatures, when being cooled (a) and heated (b)}
\end{figure}
\section{Conclusions}
1.~The structural transition at $T_{S1}$ from the
$Pmm~(a_{p}\times 2a_{p}\times 2a_{p})$ to $Pmma(2a_{p}\times 2a_{p}\times2a_{p})$
phase is observed in three isomorphic materials
$R$BaCo$_{2}$O$_{5+\delta }$, ($R$~=~Gd,~Tb$_{0.9}$Dy$_{0.1}$,~Dy) with $\delta=0.50(2)$
and the well ordered Co$^{3+}$ sites in pyramidal and octahedral
coordination. This phase transition is very likely to be a general
phenomenon, at least for the $R$ ions from the middle of the rare-earth period.

2.~The temperature of the structural phase transition $T_{S1}$ and the $\rho(T)$
anomaly coincide with a precision of 2~--~3~K, which gives indication to the
orbital ordering at this transition. The details of this ordering, in
principle, can be derived from the distances between the cobalt ions and
intermediate ligands.

3.~A structural phase transition of the first order to the phase with the
unit cell 2$a_{p}$~$\times $~2$a_{p}$~$\times $~4$a_{p}$ is discovered in the
magnetically ordered state. This transition occurs at $200<~T<210$~K on
heating, while on cooling it is extended from 150~K to 110~K. Unusual
magnetic properties observed in this temperature range \cite{12,13} probably may
be explained by coexistence of two phases with different crystal and
magnetic structures.

4.~The systematic extinction of the superstructure reflections in the
low-temperature phase does not yield the space group $Pcca$ that has been
postulated \cite{6} as a highest subgroup of $Pmma$ with the wave-vector $\textbf{k}=\textbf{c}_3^\ast/2$, 
an argument that can be only used in the case of the second ordertransitions.
\section{Acknowledgments}
We are grateful to D.I.Khomskii for the discussions that stimulated a part
of experiments and S.V.~Gavrilov for the help in the preparation the paper.
The work was supported by Russian Foundation for Basic Research (grants
05-02-17466-a, 06-02-81029-Bel{\_}a), Foundation for Fundamental Research of
Belarus (grants FR 06-068, F 05-129) and by Russian State Program ``Quantum
Macrophysics''.

\end{document}